\documentclass[reprint,preprintnumbers,nofootinbib,amsmath,amssymb,aps,prb]{revtex4-2}

\usepackage{booktabs}
\usepackage{graphicx}
\usepackage{hyperref}

\hypersetup{
  colorlinks=true,
  linkcolor=blue,
  citecolor=blue,
  urlcolor=black
}

\newcommand{\A}{A}
\newcommand{\B}{B}
\newcommand{\C}{C}
\newcommand{\md}{\mathrm{d}}

\newcommand{\abs}[1]{\left\lvert{#1}\right\rvert}
\newcommand{\absl}[1]{\lvert{#1}\rvert}
\newcommand{\avg}[1]{\left\langle{#1}\right\rangle}

\renewcommand{\vec}[1]{\mathbf{#1}}
\newcommand{\unitvec}[1]{\hat{\mathbf{#1}}}

\begin{document}

\title{Simulating magnetic antiskyrmions on the lattice}

\author{Juan C.~Criado}
\email{juan.c.criado@durham.ac.uk}
\affiliation{Institute for Particle Physics Phenomenology, Department of Physics, Durham University, South Road, Durham DH1 3LE, United Kingdom}

\author{Peter D.~Hatton}
\email{p.d.hatton@durham.ac.uk}
\affiliation{Centre for Materials Physics, Department of Physics, Durham University, South Road, Durham DH1 3LE, United Kingdom}
  
\author{Sebastian Schenk}
\email{sebastian.schenk@durham.ac.uk}
\affiliation{Institute for Particle Physics Phenomenology, Department of Physics, Durham University, South Road, Durham DH1 3LE, United Kingdom}

\author{Michael Spannowsky}
\email{michael.spannowsky@durham.ac.uk}
\affiliation{Institute for Particle Physics Phenomenology, Department of Physics, Durham University, South Road, Durham DH1 3LE, United Kingdom}

\author{Luke A.~Turnbull}
\email{l.a.turnbull@durham.ac.uk}
\affiliation{Centre for Materials Physics, Department of Physics, Durham University, South Road, Durham DH1 3LE, United Kingdom}

\preprint{IPPP/21/34}

\begin{abstract}
Magnetic skyrmions are topologically protected spin structures that naturally emerge in magnetic materials.
While a vast amount of effort has gone into the study of their properties, their counterpart of opposite topological charge, the antiskyrmion, has not received as much attention.
We aim to close this gap by deploying Monte Carlo simulations of spin-lattice systems in order to investigate which interactions support antiskyrmions, as well as skyrmions of Bloch and N\'eel type.
We find that a suitable combination of ferromagnetic exchange and Dzyaloshinskii-Moriya (DM) interactions is able to stabilize all three types.
Considering a three-dimensional spin lattice model, we provide a finite-temperature phase diagram featuring a stable antiskyrmion lattice phase for a large range of temperatures.
In addition, we also shed light on the creation and annihilation processes of these antiskyrmion tubes and study the effects of the DM interaction strength on their typical size.
\end{abstract}

\maketitle

\section{Introduction}

Magnetic materials with chiral Dzyaloshinskii-Moriya (DM) interactions~\cite{Dzyaloshinskii:1958,Moriya:1960zz} have been shown to support the emergence of topologically protected particle-like spin textures, known as (magnetic) skyrmions~\cite{Skyrme:1962vh,Bogdanov:1989}.
In general, skyrmions are characterized by their topological charge.
For instance, in two-dimensional systems, a magnetic skyrmion of unit charge can be thought of as a localized spin structure where the local magnetization points into every possible direction at least once.
Consequently, the skyrmion will typically appear as a sharply localized core in which the magnetization is inverted with respect to the surrounding spins.

The experimental discovery of magnetic skyrmions represents an exceptional opportunity to study topological solitons that are realized in nature (for a brief overview, see~\cite{Bogdanov:2020}).
For instance, the magnetic material MnSi has been found to exhibit a thermodynamical phase featuring a hexagonal lattice of stable skyrmion tubes~\cite{Muhlbauer:2009}.
This phase has also been established in a range of other materials~\cite{Munzer:2009var,yu2010real,yu2011near,tokunaga2015new,woo2016observation,fujima2017thermodynamically}.
These systems are, therefore, promising candidates for investigating the fundamental nature of magnetic skyrmions.
In addition, the development of techniques for their manipulation may even allow for applications in the field of spintronics, including magnetic data storage technology~\cite{Fert:2013}, racetrack memory~\cite{tomasello2014strategy}, artificial synapses for neuromorphic computing~\cite{song2020skyrmion}, reservoir computing~\cite{pinna2020reservoir}, and reshuffling for signal decorrelation in probabilistic computing~\cite{zazvorka2019thermal}.

Theoretical advances have further supplemented the rapid experimental developments.
The latter have focused on understanding the mechanism by which magnetic skyrmions may be formed or rendered stable (see, e.g.,~\cite{cortes2017thermal,buttner2018theory,birch2021topological}).
In particular, Monte Carlo (MC) simulations have been proven to be a powerful tool in this endeavor~\cite{do2009skyrmions,yu2010real}.\footnote{More generally, MC simulations have been successfully used to study topological solitons in quantum field theories~\cite{Brendel:2009mp,Schenk:2020lea}.}
For the example of chiral magnets, MC techniques enabled demonstrating that the combination of ferromagnetic exchange and DM interactions is sufficient to reproduce the typical finite-temperature phase diagram, featuring a stable skyrmion pocket~\cite{Buhrandt:2013uma}.
These findings immediately pose the question whether the stabilization of other topological solitons is possible in similar magnetic materials.
In particular, in this context, the counterpart of magnetic skyrmions of opposite topological charge, known as antiskyrmions, have been paid little attention to, with a few notable exceptions~\cite{koshibae2016theory,hoffmann2017antiskyrmions,huang2017stabilization,camosi2018micromagnetics,kovalev2018skyrmions,bottcher2018b,jena2020elliptical}.
While their theoretical description appears to be reasonably close to ordinary skyrmions, they have eluded any experimental evidence in magnetic materials in which DM interactions are dominant.
Nevertheless, they have been observed in systems that also support skyrmions and (topologically trivial) magnetic bubbles~\cite{peng2020controlled}, indicating the presence of interactions other than DM.
In addition, both skyrmions and antiskyrmions are also supported in frustrated magnets~\cite{okubo2012multiple,leonov2015multiply,lin2016ginzburg,Sutcliffe:2017aro}.
In practice, potential applications of antiskyrmions may even go beyond those of skyrmions.
For example, they are expected to feature an anisotropic Hall angle, in theory providing for greater control over their manipulation in spintronics~\cite{huang2017stabilization}.

Our work is supposed to close this gap by shedding light on the stabilization of magnetic antiskyrmions in chiral magnetic materials, exclusively featuring DM interactions.
Using MC simulations of a simple spin-lattice system describing the local interactions of a chiral magnet, we explore the thermodynamical phases of the material.
We show that antiskyrmions are indeed stabilized in a large region of the parameter space, given a suitable DM interaction strength.
Along these lines, we also shed light on their creation and annihilation processes.
We also confirm the existence of magnetic skyrmions of Bloch and N\'eel type, depending on the precise form of the DM interaction.
We hope that this survey will provide crucial guidance for future experiments in the search for antiskyrmions in magnetic materials.

This work is organized as follows.
In Section~\ref{sec:lattice-model}, we introduce the Hamiltonian lattice model that we use to describe a chiral magnet.
In Section~\ref{ref:monte-carlo}, we briefly summarize the MC method we use to obtain the thermodynamical phases containing (anti)skyrmions.
As a brief glimpse, we display example field configurations in the topological phases of different materials.
In Section~\ref{ref:phase-diagram}, we present the first finite-temperature phase diagram for the type of material that supports antiskyrmions.
We further investigate hysteresis effects that the system is subject to, effectively leading to deformations of the phase diagram.
In addition, in Section~\ref{sec:K}, we study the effects of variations in the DM interaction strength.
We find that it has a minimal antiskyrmion-stabilizing value and confirm that it controls the size of antiskyrmions.
Finally, we summarize our results in Section~\ref{sec:conclusions}.

\section{From the continuum to the lattice model}
\label{sec:lattice-model}

In this work, we are interested in the stabilization of skyrmion phases in three-dimensional bulk chiral magnets.
As a simple description, we evolve our discussion around a coarse-grained Hamiltonian where we describe the local magnetization as a continuous vector field $\vec{M}$ with constant norm, $M = \absl{\vec{M}}$.
Our model of a chiral magnet, therefore, takes the form~\cite{bak1980theory}
\begin{equation}
 H
 =
 \int \md^3r
 \left[
   \frac{J}{2} \left(\nabla \vec{M}\right)^2
  + K \operatorname{DM}(\vec{M})
  - \vec{B} \cdot \vec{M}
 \right] \, ,
 \label{eq:hamiltonian}
\end{equation}
where the coefficients $J$ and $K$ of the ferromagnetic exchange and DM interaction are free parameters of the model.
Furthermore, $\vec{B}$ denotes an external magnetic field, which we choose to point in the $z$-direction, $\vec{B} = B \unitvec{z}$.
From first principles, we are agnostic about the precise form of the DM interaction.
In practice, it will depend on the symmetries of the system or, equivalently, on the crystal structure of the material in the microscopic description.
In fact, all possibilities of the latter can be classified, as illustrated in Table~\ref{tab:DM}.
Here, we define three types of DM interaction (which we label \A, \B\ and \C) and provide the corresponding point group of the crystal.\footnote{For an overview and discussion of the possible DM interactions, see, e.g.,~\cite{Gobel:2020mqd}.}

\begin{table*}[t]
 \centering
 \begin{tabular}{cccc}
  \toprule
  Label
  & Point group
  & $\operatorname{DM}(\vec{M})$
  & $\operatorname{DM}_d(\vec{S})$
  \\
  \midrule
  \A
  & $T$ or $O$
  & $\mathbf{M} \cdot (\nabla \times \mathbf{M})$
  & $\mathbf{S_r} \cdot \left(
   \mathbf{S}_{\mathbf{r} + \hat{\mathbf{x}}} \times \hat{\mathbf{x}}
   + \mathbf{S}_{\mathbf{r} + \hat{\mathbf{y}}} \times \hat{\mathbf{y}}
   + \mathbf{S}_{\mathbf{r} + \hat{\mathbf{z}}} \times \hat{\mathbf{z}}
   \right)
   $
  \\
  \midrule
  \B
  & $C_{nv}$
  & $\mathbf{M} \cdot \nabla M_3 - M_3 \nabla \cdot \mathbf {M}$
  & $
   \begin{array}{c}
    (\mathbf{S}_{\mathbf{r}})_1 \left[
    (\mathbf{S}_{\mathbf{r} + \hat{\mathbf{z}}})_2
    - (\mathbf{S}_{\mathbf{r} + \hat{\mathbf{y}}})_3
    \right]
    \\
    + (\mathbf{S}_{\mathbf{r}})_2 \left[
    (\mathbf{S}_{\mathbf{r} + \hat{\mathbf{x}}})_3
    - (\mathbf{S}_{\mathbf{r} + \hat{\mathbf{z}}})_1
    \right]
    \\
    + (\mathbf{S}_{\mathbf{r}})_3 \left[
    (\mathbf{S}_{\mathbf{r} + \hat{\mathbf{y}}})_1
    - (\mathbf{S}_{\mathbf{r} + \hat{\mathbf{x}}})_2
    \right]
   \end{array}
   $
  \\
  \midrule
  \C
  & $D_{2d}$
  & $\mathbf{M} \cdot (\partial_x \mathbf{M} \times \hat{\mathbf{x}} - \partial_y \mathbf{M} \times \hat{\mathbf{y}})$
  &
   $
   \begin{array}{c}
    (\mathbf{S}_{\mathbf{r}})_2 (\mathbf{S}_{\mathbf{r} + \hat{\mathbf{y}}})_3
    - (\mathbf{S}_{\mathbf{r}})_3 (\mathbf{S}_{\mathbf{r} + \hat{\mathbf{y}}})_2
    \\
    - (\mathbf{S}_{\mathbf{r}})_3 (\mathbf{S}_{\mathbf{r} + \hat{\mathbf{x}}})_1
    + (\mathbf{S}_{\mathbf{r}})_1 (\mathbf{S}_{\mathbf{r} + \hat{\mathbf{x}}})_3
   \end{array}
   $
  \\
  \bottomrule
 \end{tabular}
 \caption{Possible DM interactions parametrized in a continuum (DM) or lattice (DM$_d$) Hamiltonian formulation. The different point groups correspond to different crystal symmetries, thereby allowing for a distinct DM interaction. In reality, these are given by the crystal structure of the magnetic material.}
 \label{tab:DM}
\end{table*}

We note that, in the above Hamiltonian, we have neglected other types of (possibly long-ranged) interactions that may be present, such as dipolar or uniaxial couplings (see, e.g.,~\cite{de1980dynamic, garel1982phase}).
A more realistic treatment of a chiral magnet would need to include the latter.
However, it has been demonstrated in lattice simulations that the purely local Hamiltonian~\eqref{eq:hamiltonian} with a DM interaction of type \A\ is able to stabilize magnetic Bloch skyrmions~\cite{Buhrandt:2013uma}.
We will show that N\'eel skyrmions and antiskyrmions can also be stabilized using this Hamiltonian, with DM interactions of type \B\ and \C, respectively.
Indeed, in our setup, there is a family of skyrmion configurations related to each other by global rotations around the $z$-axis.
Besides their topological charge, skyrmions are also characterized by the local magnetization in the $xy$-plane.
For instance, the configuration where the latter always points in the radial direction of the topological defect is known as a N\'eel skyrmion.
In contrast, for a Bloch skyrmion the in-plane magnetization is perpendicular to the radial direction.
Strictly speaking, this amounts to a global rotation by an angle of $\pm \pi / 2$, also called helicity.
For a more detailed discussion, we refer the reader to, e.g.,~\cite{Gobel:2020mqd}.

In general, topologically non-trivial configurations of the magnetization field can be characterized by means of their topological charge.
Although the latter is not always conserved, it can still be useful in systems with translation symmetry along a particular direction.
In the present case, choosing the translation-invariant direction to be the $z$-axis, we define $Q$ as (see, e.g.,~\cite{Manton:2004tk})
\begin{equation}
 Q = \frac{1}{4\pi M^3} \int \md x \md y \,
 \vec{M} \cdot \left(\partial_x \vec{M} \times \partial_y \vec{M}\right) \, .
\end{equation}
In our example, field configurations with $Q = -1$ are called skyrmions, while those with $Q = +1$ are antiskyrmions.\footnote{In principle, chiral magnets can host topological solitons of arbitrary charge~\cite{rybakov2019chiral,Foster:2019rbd,Kuchkin:2020bkg}.}
Intuitively, for a given configuration of unit charge, the local magnetization points into every possible direction at least once.
In other words, $Q$ may also be coined the antiskyrmion number (i.e.~the number of antiskyrmions inside a given volume).
Continuous deformations of the magnetization field localized in some regions cannot change the value of $Q$, as long as the field's value around this region is kept fixed.
This means that configurations with different $Q$ are topologically protected from continuously evolving into each other, and in particular from unwinding into the trivial one, $Q=0$.
However, carefully note that, in general, topological stability does not necessarily imply energetic stability.
For instance, topological sectors of different charges are often separated by barriers of finite energy (see, e.g.,~\cite{cortes2017thermal}).
We further remark that strictly speaking, $Q$ counts the winding of the magnetization field around its target space, corresponding to a sphere $S^2$, as the domain of the field is traversed.
For $Q$ to be a topological invariant (according to homotopy theory), it has to classify maps between spheres.
In the present case, however, the domain of the magnetization is not a sphere as we do not conformally compactify the projection of the underlying Euclidean space $\mathbb{R}^2$.
Therefore, homotopy arguments do not apply rigorously in our scenario.
Nevertheless, the total charge $Q$ can count the number of (anti)skyrmions inside a given lattice volume, thereby proving useful for characterizing the topological phases of magnetic materials.

\subsection*{The lattice Hamiltonian of interacting spins}

Let us now turn to the Hamiltonian formulation of the chiral magnet in more detail.
To systematically explore the thermodynamical phases of this system via MC simulations, we discretize it by considering a lattice of interacting spins.
The associated Hamiltonian of the theory on a cubic lattice with uniform lattice spacing $a$ reads
\begin{equation}
\begin{split}
 H_d
 =
 - \sum_{\vec{r}} \Big[
 &
  \tilde{J} \; \mathbf{S}_{\vec{r}} \cdot \left(
   \mathbf{S}_{\vec{r} + \unitvec{x}}
   + \mathbf{S}_{\vec{r} + \unitvec{y}}
   + \mathbf{S}_{\vec{r} + \unitvec{z}}
  \right)
 \\
 &
 	+ \tilde{K} \operatorname{DM}_d \left( \vec{S}_{\vec{r}} \right)
  + \tilde{B} \cdot \left(\vec{S}_{\vec{r}}\right)_z
 \Big] \, ,
\end{split}
\label{eq:discrete-hamiltonian}
\end{equation}
where we have defined the lattice couplings $\tilde{J} = J M^2 a$, $\tilde{K} = K M^2 a^2$ and $\tilde{B} = B M a^3$, as well as the classical spin variable $\vec{S} = \vec{M} / M$. 
The discretized counterpart of each type of DM interaction on the lattice is displayed in the rightmost column of Table~\ref{tab:DM}.
As the degrees of freedom are now given in terms of (normalized) classical spins, the system is suitable to be studied using MC techniques. 
For convenience, we also define the (discrete) topological charge on the spin-lattice,
\begin{equation}
 Q_d = \frac{1}{4 \pi} \sum_{\vec{r}}
 \vec{S}_{\vec{r}} \cdot (\vec{S}_{\vec{r} + \unitvec{x}} \times \vec{S}_{\vec{r} + \unitvec{y}}) \, .
\label{eq:top_charge_discrete}
\end{equation}
Although, strictly speaking, the topological arguments we presented in the continuum case cannot be applied to the discrete one, $Q_d$ approximates $Q$ very accurately in the limit of small lattice spacing.
In this regime, $Q_d$ captures all essential properties of the topological charge reasonably well.

Before we continue, we also remark that the discretization of the field degrees of freedom on a finite lattice generically introduces inaccuracies.
In principle, one is free to choose the above discretization or any other, since the large-distance behaviour of the system should be independent of the microscopic description.
However, typically, this is only true in the limit of a small lattice spacing, $a \to 0$, where the continuum theory is recovered.
The continuum limit of physical observables should therefore be taken with caution.
For instance, in practice, for finite, non-zero lattice spacing, spurious anisotropies may appear~\cite{Buhrandt:2013uma}.
These are further deteriorated due to the finite volume of the spin-lattice.
Closely following~\cite{Buhrandt:2013uma}, here, we aim to correct for these by introducing counter terms associated with next-to-nearest-neighbour couplings.
The latter can lead to a partial cancellation of the anisotropies, as seen in momentum space.
Let $H_d^{\prime}$ be the next-to-nearest neighbor Hamiltonian, its interactions being the ferromagnetic exchange and DM terms, with coefficients $\tilde{J}^{\prime}$ and $\tilde{K}^{\prime}$, respectively.
In momentum space, the coefficient of the ferromagnetic exchange interaction of the total Hamiltonian, $H_d + H_d^{\prime}$, is given by
\begin{equation}
 \alpha \left(\vec{q}\right) = \tilde{J} \cos\left(a \abs{\vec{q}}\right) + \tilde{J}^{\prime} \cos\left(2 a \abs{\vec{q}}\right) \, .
\end{equation}
Thus, examining the series expansion in the lattice spacing, higher-order terms in $a$ contain powers of the momentum $\absl{\vec{q}}$ greater than two. At the same time, the only contribution to the continuum interaction is $\absl{\vec{q}}^2$, because of the two derivatives.
We therefore set $\tilde{J}^{\prime} = -\tilde{J} / 16$ in order to cancel the first non-trivial correction, corresponding to the $\absl{\vec{q}}^4$ term.
A similar procedure can be applied to the DM interaction term.
The coefficient of the latter has a more complex tensor structure, but the momentum-dependence can be schematically summarized in the following vector of coefficients,
\begin{equation}
 \beta_i \left(\vec{q}\right) = \tilde{K} \sin(a q_i) + \tilde{K}^{\prime} \sin(2 a q_i) \, .
\end{equation}
In this case, the continuum interaction is proportional to the linear contribution $\absl{\vec{q}}$, and we use $\tilde{K}^{\prime} = -\tilde{K} / 8$ to cancel its first correction proportional to $\absl{\vec{q}}^3$.
For a more detailed discussion of the cancellation of spurious anisotropies due to discretization effects, we refer the reader to~\cite{Buhrandt:2013uma}.

\section{Thermodynamical phases from Monte Carlo simulations}
\label{ref:monte-carlo}

To explore the thermodynamical phases of the magnetic material, the main object of interest is the thermal expectation value of the local magnetization.
In the previous section, we have already identified the corresponding degrees of freedom with a lattice of interacting classical spins.
Therefore, finding the thermal expectation value of the magnetization requires us to investigate the possible spin configurations at any given temperature, as can be seen from the path integral,
\begin{equation}
	\avg{\vec{S}} = \frac{1}{\mathcal{Z}} \int \mathcal{D} \vec{S} \, \vec{S} \exp \left(- \frac{H_d}{k_B T}\right) \, .
\end{equation}
Here, $T$ is the temperature, and $k_B$ is the Boltzmann constant.
Furthermore, $\mathcal{Z}$ denotes the partition function of the theory.
Naively, this expression implies that the thermal expectation value is dominated by spin configurations that minimize the Hamiltonian at any given temperature.
To evaluate $\avg{\vec{S}}$, we, therefore, need to find configurations of minimal energy.
This is a high-dimensional optimization problem well-suited for deploying MC techniques.
In particular, to determine the dominant spin configurations, we use a simulated annealing method, as we will briefly describe below.

To explore the spin configuration space of the theory, within our MC approach, we have to sample spin-lattice configurations following the Boltzmann distribution $\exp \left(-H_d /(k_B T)\right) / \mathcal{Z}$.
Formally, however, this configuration space is infinite-dimensional, clearly obstructing the \emph{ad hoc} generation of samples.
That means it is computationally not feasible to randomly construct spin configurations and \emph{a posteriori} determine their associated weight inside the path integral.
Instead, we want to use a simulated annealing process through the well-known Metropolis-Hastings algorithm~\cite{Metropolis:1953am,Hastings:1970aa}, which will robustly generate the desired distribution of spin samples as follows.
First, we randomly initialize a lattice of arbitrary spins.
Then, consecutively, each spin of the lattice is probed by replacing it randomly.
While probing each spin, the change in energy, $\Delta H_d$, is measured and the generated spin variable is accepted with probability $\exp \left( - \Delta H_d / (k_B T)\right)$.
In principle, the system will slowly converge towards a spin configuration of lower and lower energy, in turn dominating the thermal expectation value of the local magnetization.

The convergence of this process towards the spin configurations of minimal energy crucially depends on the system's temperature.
Ideally, to remove any bias from initial conditions, we, therefore, initialize the procedure in the high-temperature regime, $T \to \infty$, where virtually any spin replacement is accepted.
We then cautiously cool down the system by slowly lowering $T$ to the desired value that we want to probe.
This step is coined the \emph{thermalization} process, where the system adjusts to the new temperature.
At any given temperature, we can then record the desired number of sample spin configurations and take their average to obtain $\avg{\vec{S}}$.
The precise way in which we lower the temperature, we will call a \emph{schedule}.
Quite remarkably, here, the temperature is a physical parameter that, at the same time, controls the thermal fluctuations of the theory when moving through configuration space.

It is also worth noting that the path integral, and therefore our MC simulations, remain invariant if both the Hamiltonian $H_d$ of the theory and the temperature $T$ are multiplied by a numerical factor.
Thus, a simultaneous rescaling of $\tilde{J}$, $\tilde{K}$, $\tilde{B}$ and $T$ will leave any observable unchanged.
We, therefore, normalize the latter with respect to the kinetic term by defining the ratios
\begin{equation}
 \hat{B} = \frac{\tilde{B}}{\tilde{J}} \, ,
 \quad
 \hat{K} = \frac{\tilde{K}}{\tilde{J}} \, ,
 \quad
 \hat{T} = k_B \frac{\tilde{T}}{\tilde{J}} \, .
\end{equation}
The above dimensionless quantities represent the only free parameters of the simulation.
Therefore, for the rest of this work, we give our results in terms of these.
In practice, we set $\tilde{J} = 1$ without loss of generality.

\begin{figure}
 \centering
 \includegraphics[width=0.875\columnwidth]{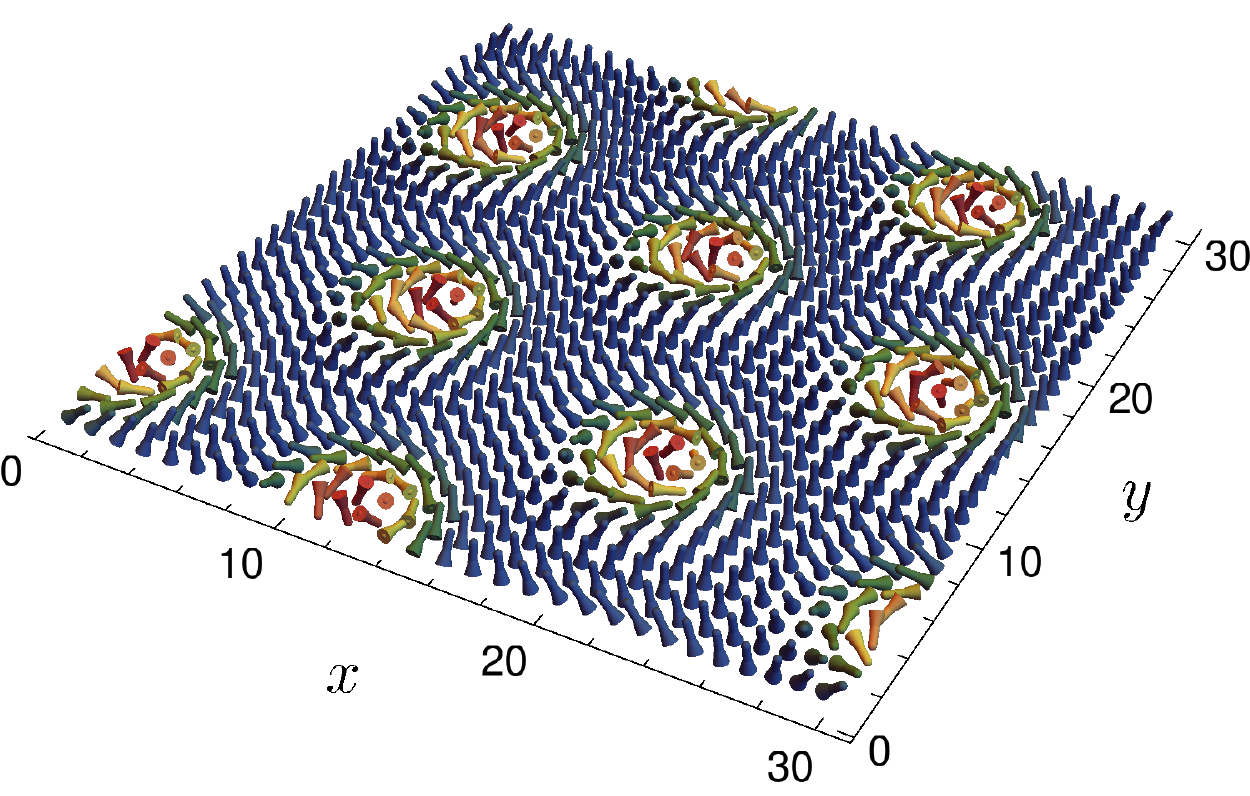}
 \includegraphics[width=0.875\columnwidth]{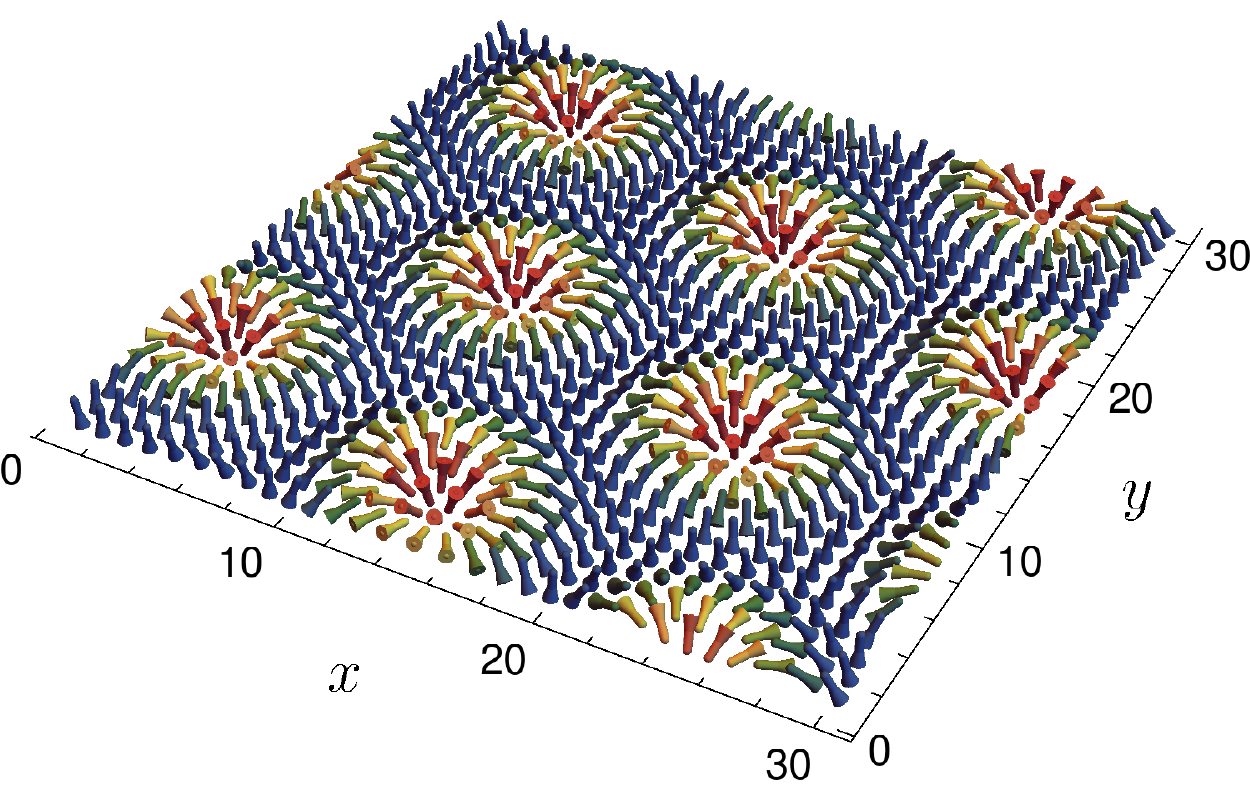}
 \includegraphics[width=0.875\columnwidth]{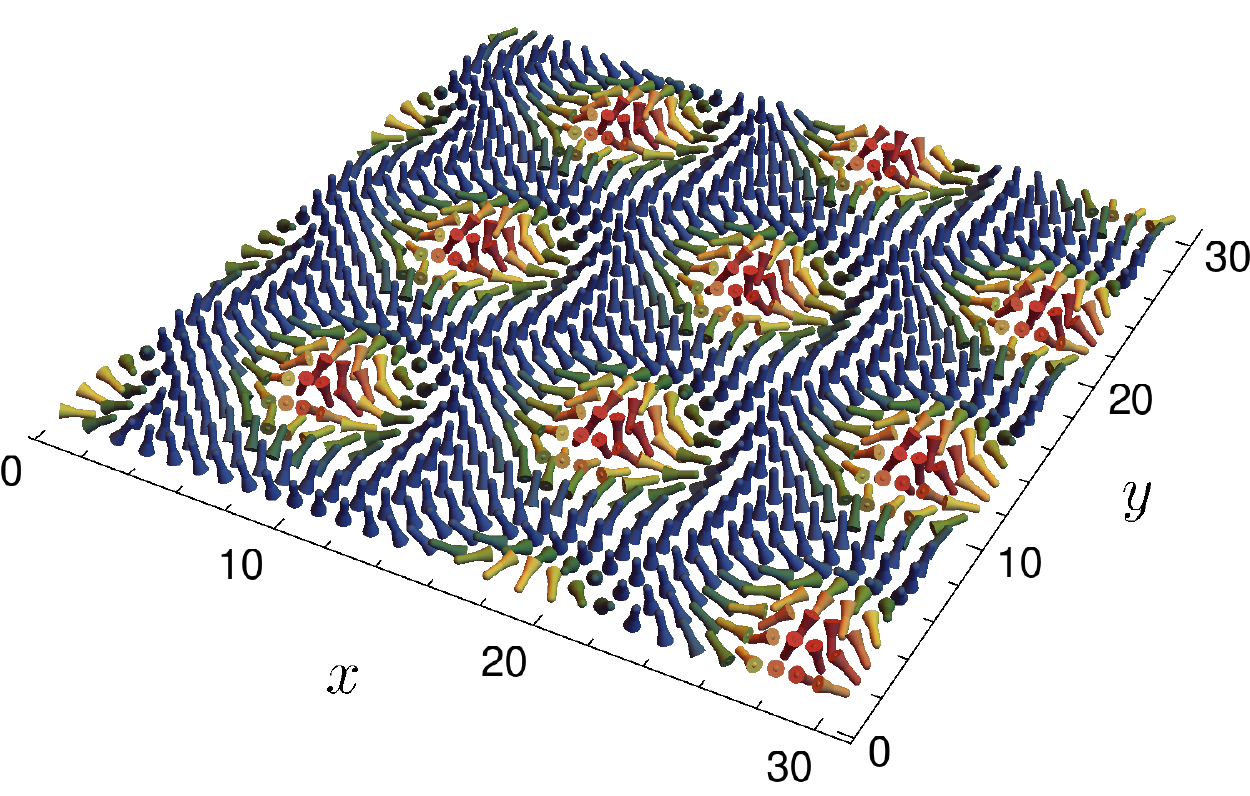}
 \caption{Slices of the thermal expectation value of the spin lattice, $\avg{\vec{S}}$, at $\hat{T}=0.80$ and $\hat{B}=0.15$ for DM interactions of type \A, \B\ and \C, corresponding to Bloch skyrmions, N\'eel skyrmions and antiskyrmions, from top to bottom. The colors denote the $z$-component of the spins, where $S_z = 1$ is shown in blue, while $S_z = -1$ is shown in red.}
 \label{fig:types}
\end{figure}

\begin{figure}
 \centering
 \includegraphics[width=0.8\columnwidth]{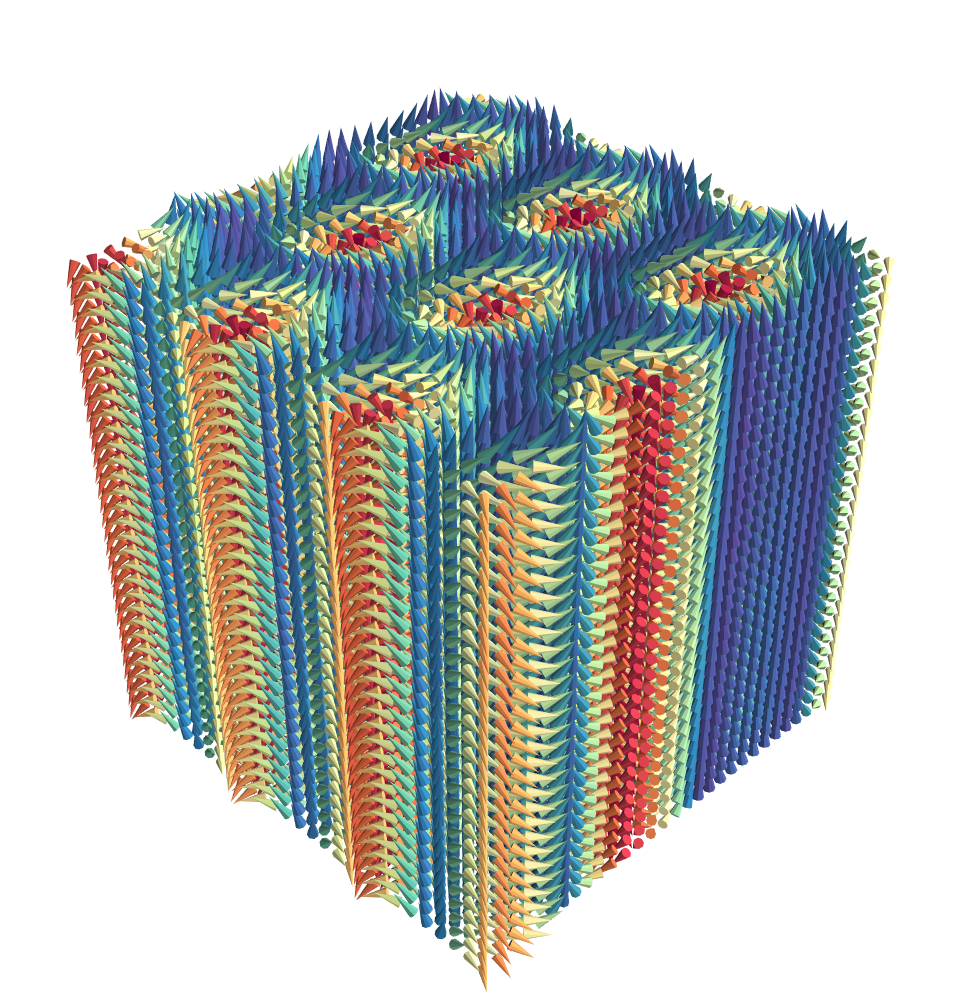}
 \includegraphics[width=0.8\columnwidth]{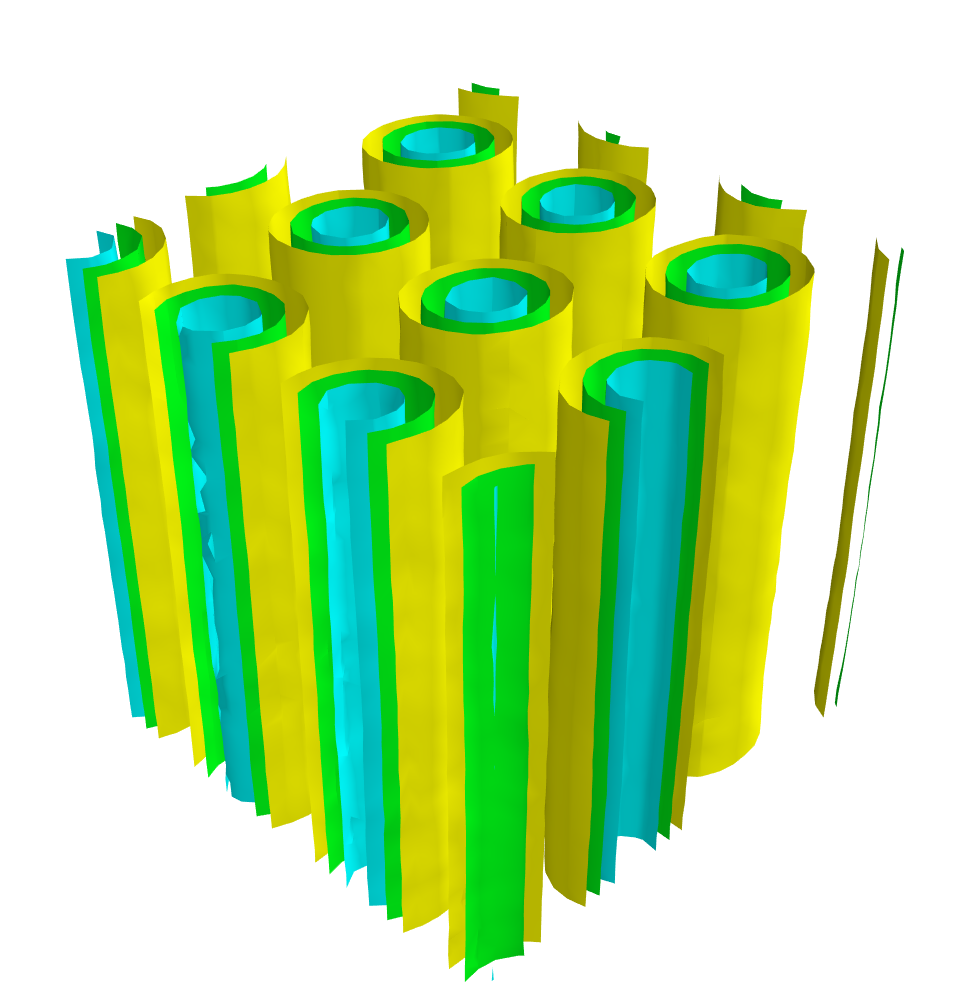}
 \caption{Three-dimensional representations of the average spin lattice configuration at $\hat{T}=0.80$ and $\hat{B}=0.15$ for DM interactions of type \C, corresponding to antiskyrmions. In the top panel, the colors denote the $z$-component of the spins. The bottom panel illustrates the contour surfaces of the $z$-component.}
 \label{fig:3D}
\end{figure}

Let us now demonstrate how to utilize MC techniques to explore the emergence of magnetic (anti)skyrmions in a chiral magnet, featuring the Hamiltonian~\eqref{eq:discrete-hamiltonian}.
Our simulation primarily uses a lattice of $30 \times 30 \times 30$ spins with periodic boundary conditions.
From a technical point of view, to drastically speed up the algorithm, we also divide the spin-lattice into non-interacting domains.
That is, in practice, following the approach presented in~\cite{romero2019performance}, the lattice is divided into three sublattices in a checkerboard pattern so that any given spin, its nearest neighbours as well as its next-to-nearest neighbours are each contained in a different sublattice.
This allows us to use a Metropolis-Hastings algorithm in which all spins belonging to the same sublattice are updated in parallel, which we implement in a GPU.
We achieve a simulation speed of about $10^9$ spin updates per second on a Tesla V100 GPU with this setup.
Then, after an annealing schedule to find the thermal ground state, we average over 2000 configurations, with 50 lattice sweeps of separation between each other, to determine $\avg{\vec{S}}$.

As a very first example, we find that a simple schedule with constant temperature $\hat{T} = 0.80$, magnetic field $\hat{B} = 0.15$ and DM interaction strength $\hat{K} = \tan \left(2 \pi / 10\right)$, for $10^5$ lattice sweeps (i.e.~each spin of the lattice is probed $10^5$ times by the MC algorithm) is able to generate skyrmions as well as antiskyrmions.
The precise type depends on the form of the DM interaction, as shown in Table~\ref{tab:DM}.
Here, type \A\ corresponds to Bloch skyrmions, type \B\ corresponds to N\'eel skyrmions, and type \C\ corresponds to antiskyrmions.
Slices of the average spin configuration obtained in each case are shown in Fig.~\ref{fig:types}.
Furthermore, Fig.~\ref{fig:3D} illustrates three-dimensional representations of the magnetic antiskyrmion configuration.
As pointed out in the previous section, we can count the number of (anti)skyrmions inside the lattice volume by their topological charge.
Although the latter is not a topological invariant in the present case, we still find that these examples exhibit a total charge of $Q_d \approx \pm 8$.\footnote{Intuitively, one could naively think of it as a topological invariant if we identify the boundary around every vortex with a point.}
We also note that it is \emph{a priori} not guaranteed that this schedule at a constant temperature yields topologically non-trivial spin configurations.
Let us, therefore, explore the thermodynamical phases of the chiral magnet in more detail in the following section.

\section{The antiskyrmion lattice phase}
\label{ref:phase-diagram}

In contrast to the previous section, we will exclusively focus on \C-type materials, supporting antiskyrmions.
To examine the thermodynamical phases of the chiral magnet, we use a more realistic annealing schedule.
In particular, we choose a schedule that mimics the most common experimental technique.
That is, we implement a zero-field cooling (ZFC) procedure by starting at a high temperature of $\hat{T} = 2$ and vanishing magnetic field, $\hat{B} = 0$.
We then exponentially decrease the temperature down to the desired one in 20 steps.
At each step, we let the system thermalize for 10,000 lattice sweeps.
We then increase the magnetic field linearly, and at each value, we average over 2000 configurations, with 50 sweeps of separation between each contiguous pair, to compute the thermal expectation value of the spin configuration, $\avg{\vec{S}}$.
The DM interaction coefficient is fixed to the value $\hat{K} = \tan (2 \pi / 10)$ throughout all simulations.

\subsection{The finite-temperature phase diagram}

\begin{figure}
 \centering
 \includegraphics[width=0.8\columnwidth]{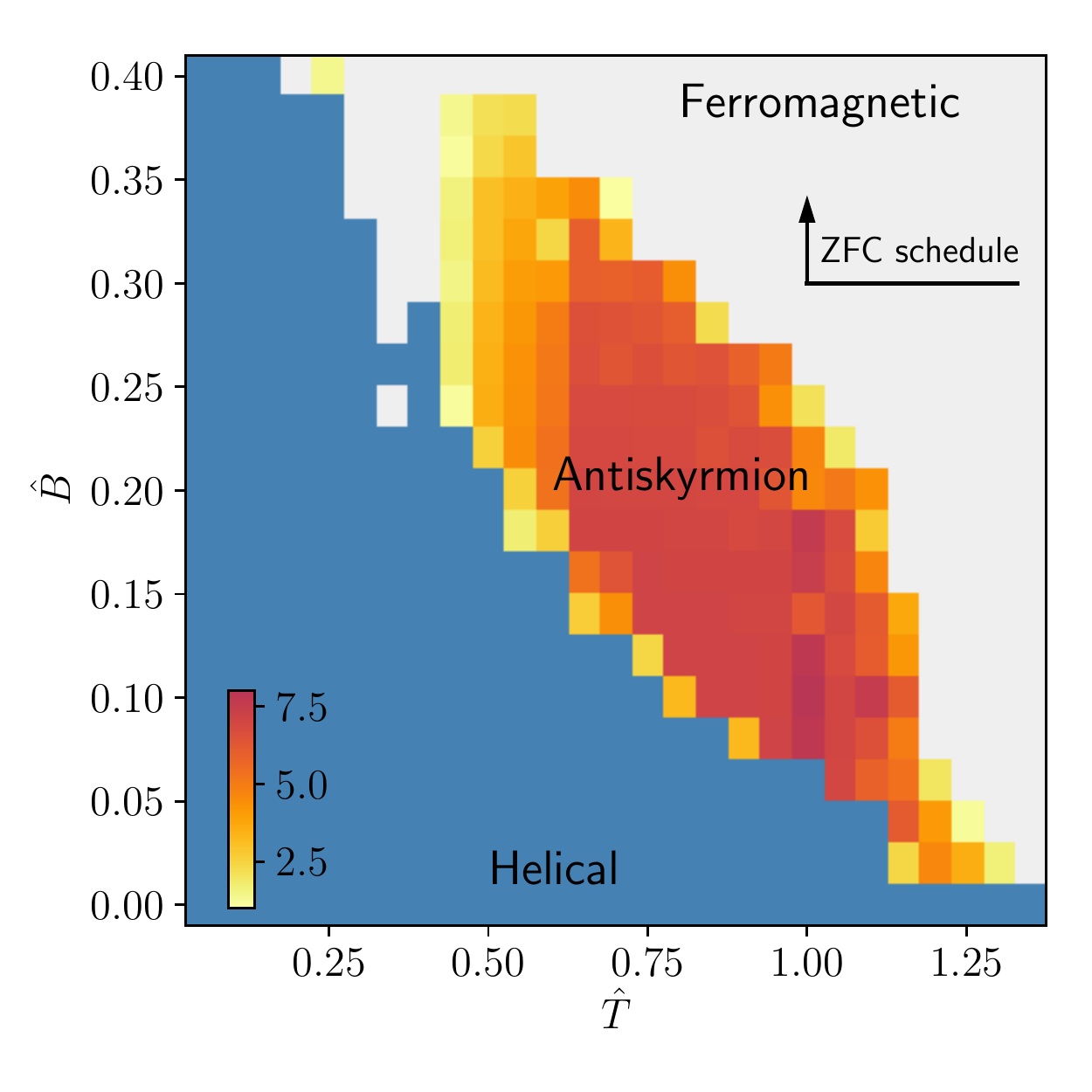}
 \caption{Finite-temperature phase diagram containing antiskyrmions for a zero-field cooling (ZFC) schedule. The schedule is schematically shown in the figure. Red regions illustrate stable antiskyrmion tubes. Here, the DM interaction coefficient is fixed to $\hat{K} = \tan \left(2 \pi / 10\right)$. The color-coding illustrates the total antiskyrmion number, $Q$.}
  \label{fig:phase-diagram}
\end{figure}

The thermodynamical phases we obtain via our MC algorithm are summarized in Fig.~\ref{fig:phase-diagram}.
The colour-coding illustrates the total antiskyrmion number, $Q$.
Therefore, in the red region, the average configuration is a hexagonal lattice of antiskyrmion tubes with cylindrical symmetry, clearly identifying the antiskyrmion phase of the material (see also Fig.~\ref{fig:3D}).
Indeed, this arrangement is correspondingly similar to what has been observed for skyrmions in materials with DM interactions of the other two types~\cite{neubauer2009topological, kezsmarki2015neel}.
A lighter red to yellow colouring is used for points that contain antiskyrmions without such a compact packing.
For illustration, we use these colours for any point for which there is at least one antiskyrmion tube present, $Q \geq 1$.
We notice an extended yet clearly bounded region in which a hexagonal lattice of antiskyrmion tubes is rendered stable.
For the points with $Q < 1$, we compute the Fourier transform of the spin configuration and count the number of intensity peaks.
Regions with one peak correspond to the ferromagnetic phase and are displayed in light grey.
The other region, shown in blue, belongs to the helical phase.

\begin{figure}
 \centering
 \includegraphics[width=0.8\columnwidth]{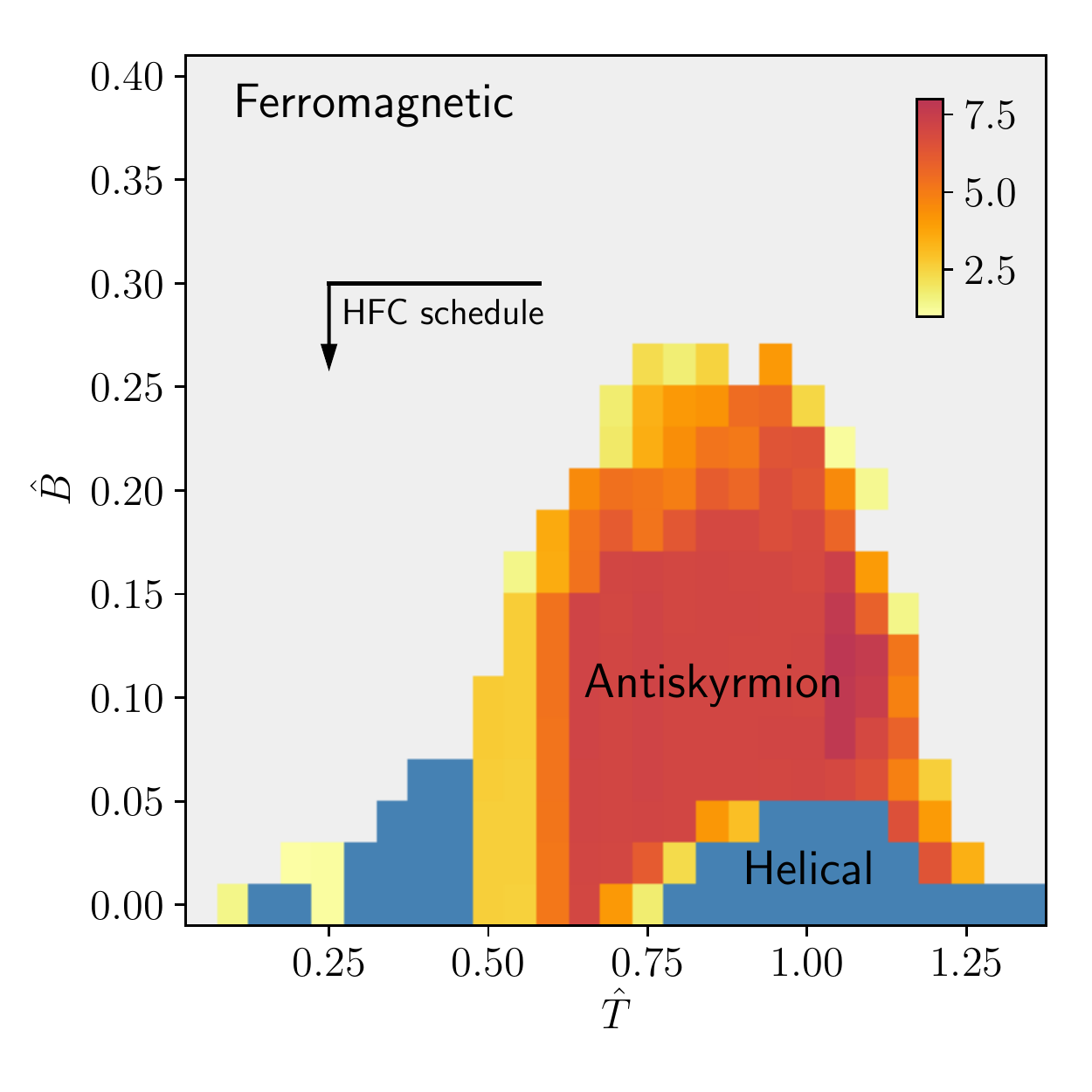}
 \includegraphics[width=0.8\columnwidth]{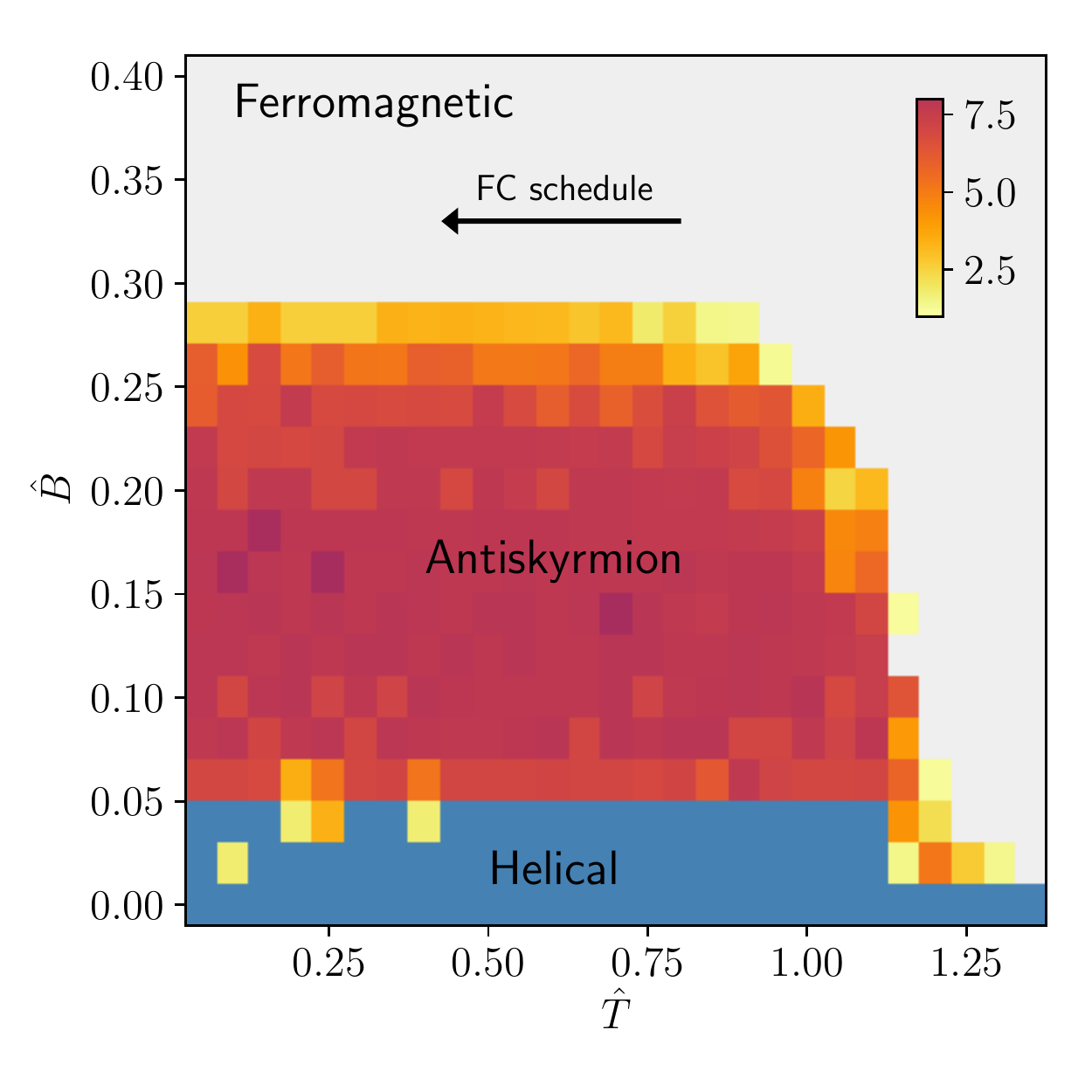}
 \caption{Finite-temperature phase diagram similar to Fig.~\ref{fig:phase-diagram} but for different annealing schedules, i.e., high-field cooling (HFC, top) and constant-field cooling (FC, bottom). Both display relatively strong hysteresis effects as we explain in the main text.}
 \label{fig:hysteresis}
\end{figure}

\begin{figure*}
 \centering
 \includegraphics[width=0.29\textwidth]{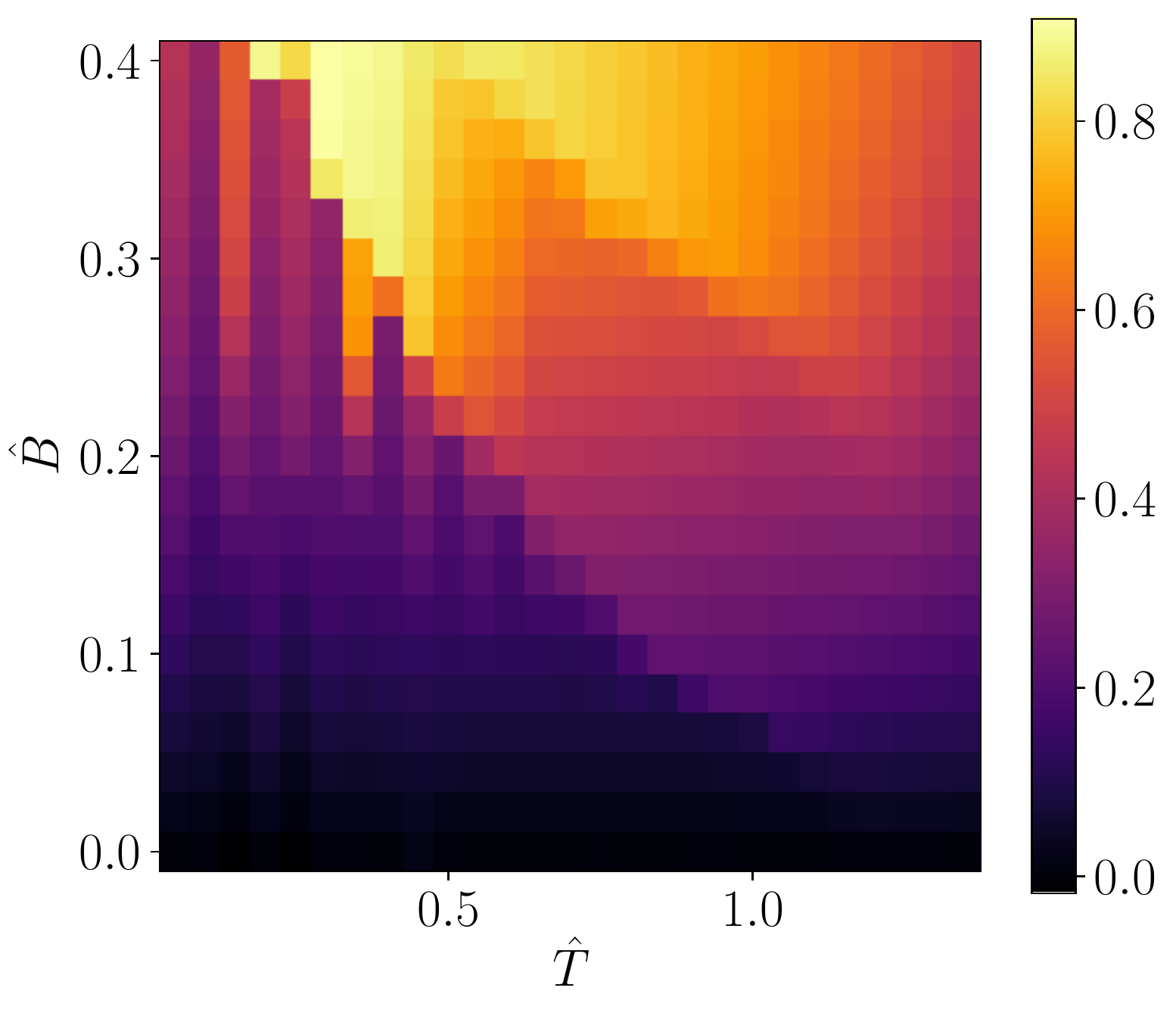}
 \hspace{0.5cm}
 \includegraphics[width=0.25\textwidth]{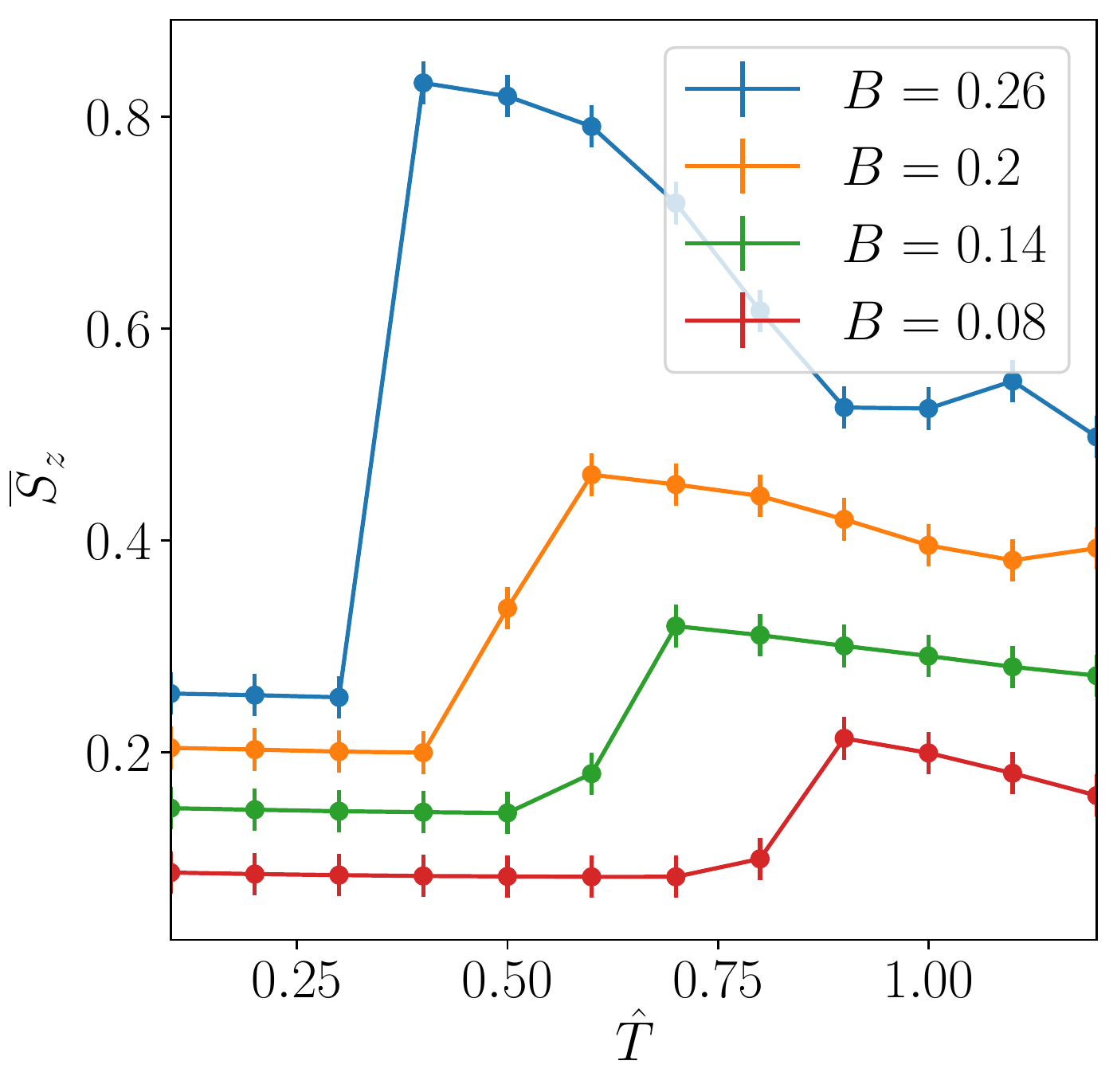}
 \hspace{0.5cm}
 \includegraphics[width=0.25\textwidth]{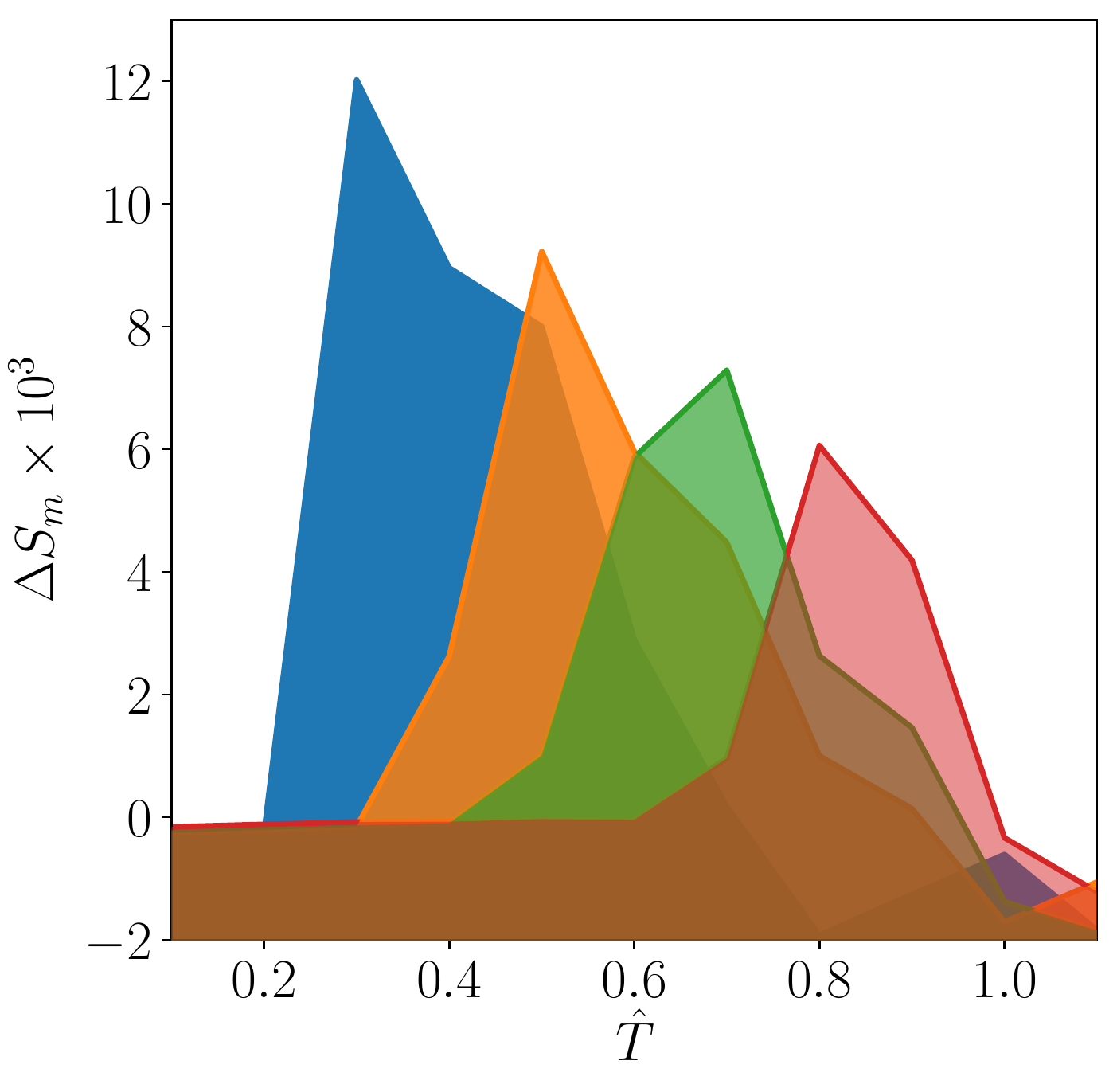}
 \caption{Dimensionless average magnetization $\bar{S}_z$ as a function of the temperature $\hat{T}$ and magnetic field $\hat{B}$ (left and center) and magnetic entropy differences $\Delta S_m$ as function of $\hat{T}$ (right). The colors in the right panel coincide with the ones in the center panel. The error bars in the center panel indicate the uncertainties introduced by the MC algorithm and may likely underestimate the total uncertainty (see main text).}
 \label{fig:magnetization}
\end{figure*}

We also note that the system is subject to strong hysteresis effects.
To study these, we use two alternative annealing schedules for our simulation, corresponding to high-field cooling (HFC) and constant-field cooling (FC), respectively.
For the HFC schedule, we follow a similar procedure as for the ZFC one, with the only difference being that the magnetic field is initialized at a relatively high value, $\hat{B} = 0.5$, and decreased linearly in 20 steps down to the desired value between the cooling and the averaging stages.
In the FC schedule, we fix the target magnetic field from the beginning and perform the exponential cooling from $\hat{T} = 2$ for each point.
Both possibilities are shown in Fig.~\ref{fig:hysteresis}.
The finite-temperature phase diagram exhibits strong hysteresis effects, manifest as deformations of the regions corresponding to each phase in $(\hat{T}, \hat{B})$ space.
For instance, the thermodynamical state of points with a higher magnetic field survives to even lower values in the HFC procedure.
A similar effect can be seen for the temperature in the FC schedule.
This may indicate that the MC algorithm is stuck in a metastable state.
This is also commonly observed in experiments, where a skyrmion lattice phase will persist metastably to low temperatures if FC is used~\cite{birch2019increased,sukhanov2020robust}.
However, in all schedules, an antiskyrmion phase around the $\hat{T} = 0.9$, $\hat{B} = 0.15$ benchmark point is consistently present, strongly supporting the stability of a hexagonal lattice of magnetic antiskyrmion tubes in this region.

Experimentally, the antiskyrmion lattice phase may be identified by tracking phase transitions in certain measurements of observables, such as the magnetization or magnetic entropy differences.
In our scenario, we translate both into their dimensionless counterparts $\bar{S}_z$ and $\Delta S_m$, respectively.
We define the former as the average of the thermal expectation value of $S_z$ over the entire lattice volume.
In addition, the latter can be formally written as
\begin{align}
 \Delta S_m\left(T, B\right) = \int_0^B \md B^{\prime} \, \left.\frac{\partial \bar{S}_z}{\partial T}\right\rvert_{B^{\prime}} \, .
\end{align}
We illustrate both observables in Fig.~\ref{fig:magnetization} as a function of the temperature (and also magnetic field).
In practice, to obtain the results shown here, we performed a ZFC schedule to drive the system into a helical state at a temperature of $\hat{T} = 0.1$, and some target value of the magnetic field.
We then increased $\hat{T}$ in steps of 0.1, while keeping $\hat{B}$ fixed.
We let the system thermalize over $10^5$ lattice sweeps at each step and finally average over 2000 configurations, with 50 sweeps in between consecutive samples.
From this procedure, $\bar{S}_z$ can be obtained immediately, while $\Delta S_m$ can be computed through a finite-differences approximation,
\begin{equation}
 \Delta S_m \approx \sum_{B^{\prime} = 0}^{B} \Delta B^{\prime} \frac{\bar{S}_z(T + \Delta T, B^{\prime}) - \bar{S}_z(T, B^{\prime})}{\Delta T} \, .
\end{equation}
We observe that with increasing temperature, all observables exhibit a sharply localized rise at a certain critical temperature.
This critical temperature characterizes the phase transition into the stable hexagonal antiskyrmion lattice phase.
In practice, the steep rise may help to experimentally identify the phase boundaries of the latter to good precision.

\begin{figure*}
 \centering
 \includegraphics[width=0.7\columnwidth]{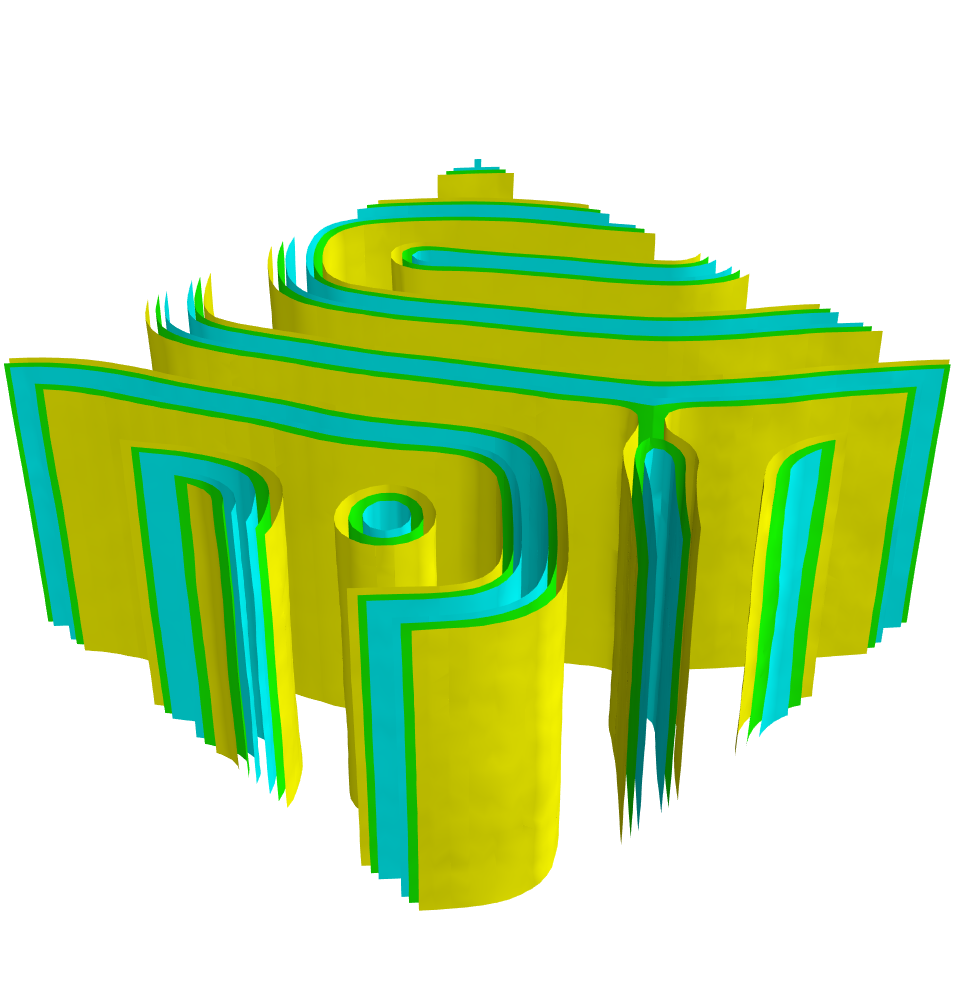}
 \hspace{2cm}
 \includegraphics[width=0.7\columnwidth]{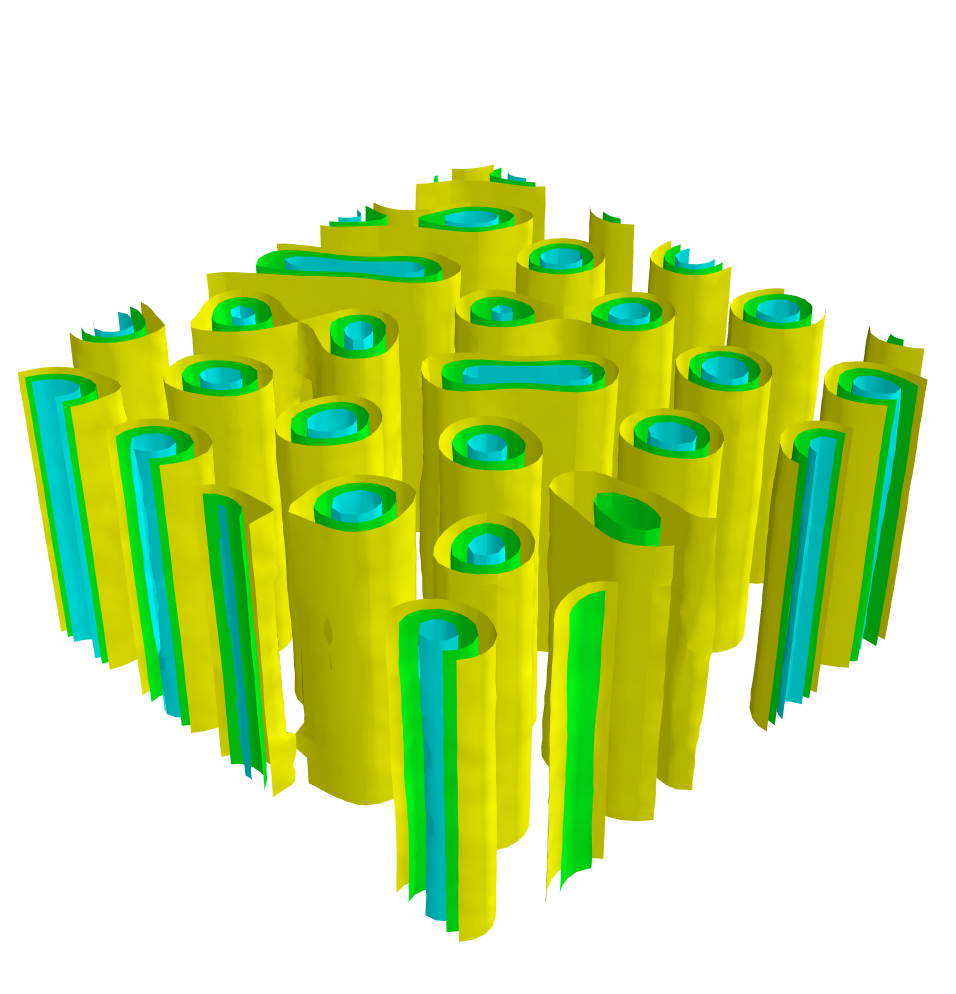}
 \caption{Contours of the $z$-component of spin configurations close to the boundary of the antiskyrmion lattice phase, obtained through a ZFC schedule. Here, we show two target temperatures, $\hat{T} = 0.45$ (left) and $\hat{T} = 0.60$ (right), while fixing the magnetic field to $\hat{B} = 0.2$. The simulations are done for a $60 \times 60 \times 30$ lattice.}
 \label{fig:other-configs}
\end{figure*}

\begin{figure*}
 \centering
 \includegraphics[width=0.7\columnwidth]{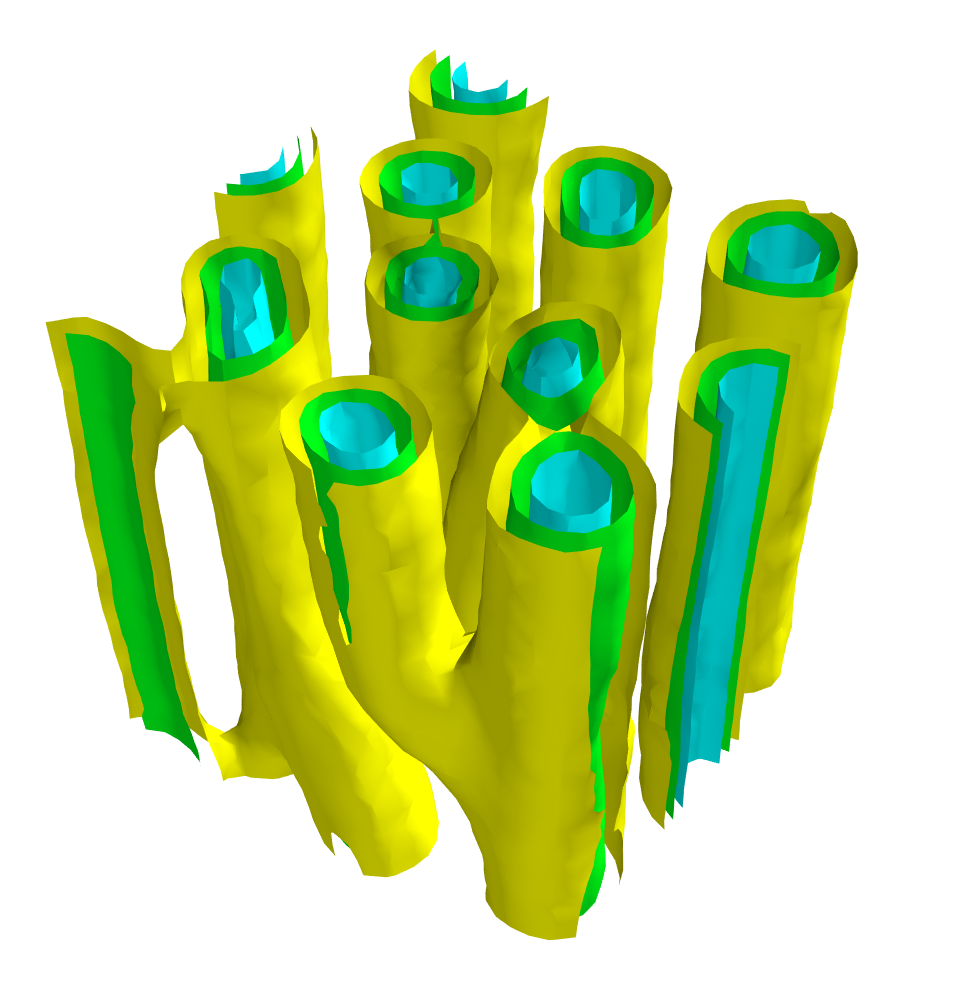}
 \hspace{1.5cm}
 \includegraphics[width=0.7\columnwidth]{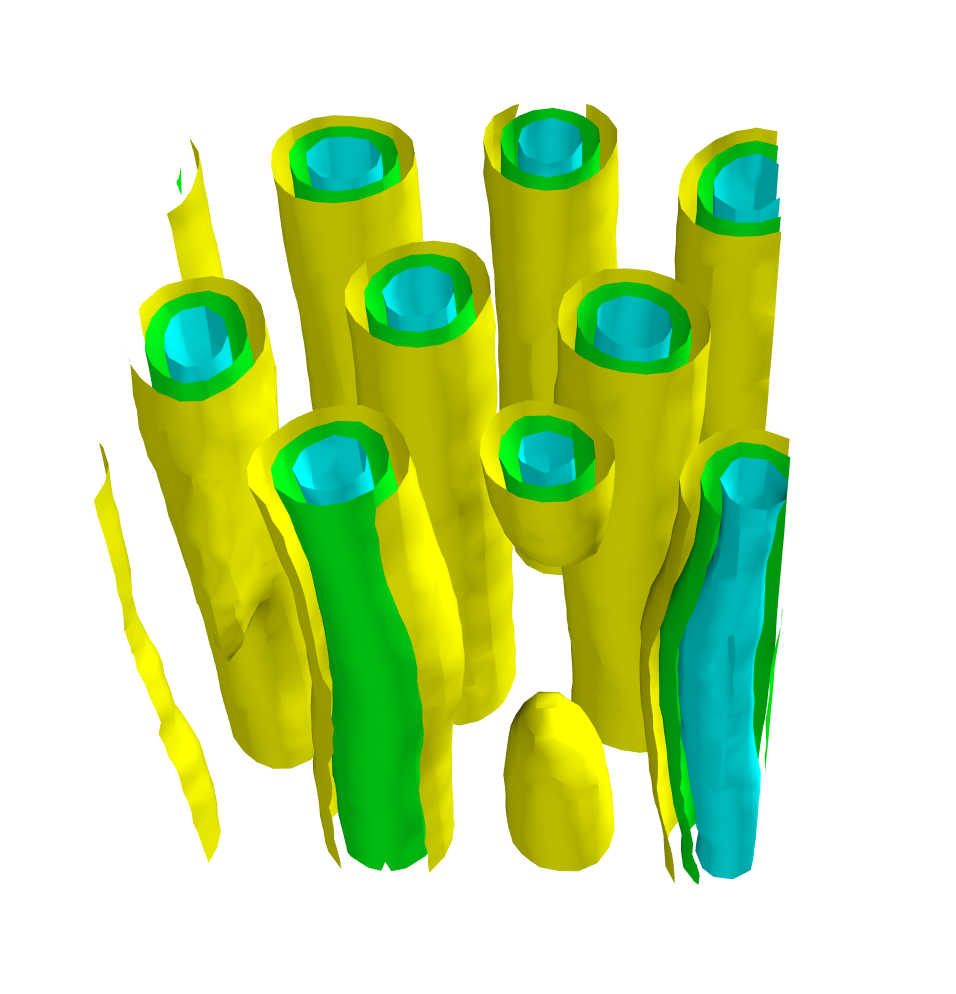}
 \caption{Snapshots of unstable spin configurations at the boundary of the antiskyrmion lattice phase. These are obtained for short MC thermalization times after a fast cooling from an initial temperature of $\hat{T} = 2$ to the target parameters $\hat{T} = 0.4$, $\hat{B} = 0.2$ (left) and $\hat{T} = 0.6$, $\hat{B} = 0.15$ (right). In the left panel several antiskyrmion tubes are branching into each other, while in the right panel an antiskyrmion tube ends abruptly. These can be identified with Bloch points that mediate the phase transition.}
 \label{fig:tubes_branching}
\end{figure*}

Let us close our discussion with a few words of caution.
Using an MC algorithm to evaluate the thermal expectation value of the spin configuration introduces errors.
Inevitably, these lead to uncertainties in the observables we have presented in this section.
For instance, formally, the error associated with the MC evaluation of the thermal expectation value $\avg{\vec{S}}$ will be of the form (see, e.g.,~\cite{Morningstar:2007zm})
\begin{equation}
	\sigma = \sqrt{\frac{\avg{\vec{S}^2} - \avg{\vec{S}}^2}{N}} \, ,
\end{equation}
where $N$ is the number of MC samples and $\vec{S}^2$ has to be understood component-wise.
For simplicity, the contribution $1 / \sqrt{N}$ is indicated in Fig.~\ref{fig:magnetization}.
However, let us remark that this form of error estimate likely underestimates the total uncertainty associated with our approach.
Most importantly, the discretization of the system on a finite lattice volume and the hysteresis effects are probably introducing even larger uncertainties.
For the latter case, we have explicitly demonstrated that, in some situations, depending on the annealing schedule, the system may get stuck in a metastable state.
Thus, it loses the capability to precisely identify phase transitions (see Fig.~\ref{fig:hysteresis}).
While this does not pose a conceptual problem, the quantitative estimates we present in this section have to be taken with some caution.
Still, the qualitative agreements with experimental observations and theoretical expectations are striking and strongly support our approach's validity.

In summary, our MC algorithm is well suited to explore the thermodynamical phases of a chiral magnet efficiently.
In particular, for materials with a \C-type DM interaction (cf.~Table~\ref{tab:DM}), we find that a hexagonal lattice of antiskyrmion tubes is stabilized for a large region of parameter space, embedded between a helical and ferromagnetic phase.
We also observe that the system is subject to strong hysteresis effects.
These depend on how phase boundaries are crossed in the annealing schedule, such that the system may end up in a metastable state, which is consistent with experimental measurements.
However, all schedules agree in a core region of phase space, strongly indicating the existence of stable magnetic antiskyrmions in the model system.

\subsection{Creation and annihilation of antiskyrmions}

In addition to establishing the thermodynamical state of the system at each point in phase space, our approach also sheds some light on the creation and annihilation of antiskyrmions.
Even though this is a dynamical process, we can gain some insights by adjusting the annealing schedule appropriately.

In a first step, we can more closely explore the boundaries of the hexagonal antiskyrmion lattice phase, where we find a mixture of different states.
As a particular example, in Fig.~\ref{fig:other-configs}, we show the thermal expectation value of the spins for fixed magnetic field, $\hat{B}= 0.2$, and two temperatures, $\hat{T} = 0.45$ and $\hat{T} = 0.60$, obtained through a ZFC schedule.
These can be understood as snapshots of the system close to the antiskyrmion phase boundary.
We find that in these configurations, the antiskyrmion tubes are partially merged into domain wall structures when crossing the phase boundary, eventually unwinding into the helical phase.
The translational symmetry along the $z$-direction is preserved throughout this process.

In addition, we can gain even more insight into the creation and annihilation of antiskyrmions when varying the rate at which the temperature is changed throughout the annealing schedule.
Naively, if the cooling rate is too high, the system may get stuck in a metastable vacuum which does not correspond to the thermal ground state.
In this way, we can force the MC algorithm to ``freeze" a specific state while crossing a phase boundary.
This allows us to obtain a snapshot of the dynamics of the process.
Physically, this metastable state has a finite lifetime, in turn depending on the temperature.
To illustrate the dynamics, we, therefore, fix the magnetic field to its target value and perform a fast cooling from initially $\hat{T} = 2$ to the target temperature, in three steps with only 200 thermalization sweeps each.
We average over 200 configurations only (separated by 50 sweeps) to capture an intermediate state of the quickly changing system before it stabilizes.
Intriguingly, in this particular example, we find that the phase transition towards the antiskyrmion phase is mediated by topological defects, similar to what has been observed for skyrmions~\cite{birch2021topological}.
These so-called Bloch points can be understood as emergent magnetic monopoles that unwind the antiskyrmion tubes, thereby annihilating them.
We illustrate snapshots of this process in Fig.~\ref{fig:tubes_branching}.
Here, we find a branching and an unwinding of the antiskyrmion tubes, which can be identified with an ending on Bloch points (see, e.g.,~\cite{birch2021topological}).
Therefore, our simulation indicates the existence of topological defects that mediate the phase transition towards the stable antiskyrmion lattice phase.
It would be interesting to investigate the dynamics of this process in more detail, similar to earlier studies, e.g., capturing the interactions of skyrmions~\cite{Foster:2019rbd,capic2020skyrmion,brearton2020magnetic,leonov2017asymmetric,Ross:2020hsw}.
We leave this for future work.

\section{Modifying the DM interaction strength}
\label{sec:K}

Previously, we have demonstrated that magnetic antiskyrmion tubes are stabilized in a large region of parameter space.
For this simulation we have fixed the (lattice) DM interaction coefficient to $\hat{K} = \tan \left(2\pi/10\right)$.
Let us now explore the antiskyrmion stability with respect to changes of this parameter.
As we have seen that antiskyrmions are consistently formed at a temperature of about $\hat{T} = 0.9$, we keep the latter value fixed to study deformations with respect to the DM interaction strength.

\begin{figure}
 \centering
 \includegraphics[width=0.8\columnwidth]{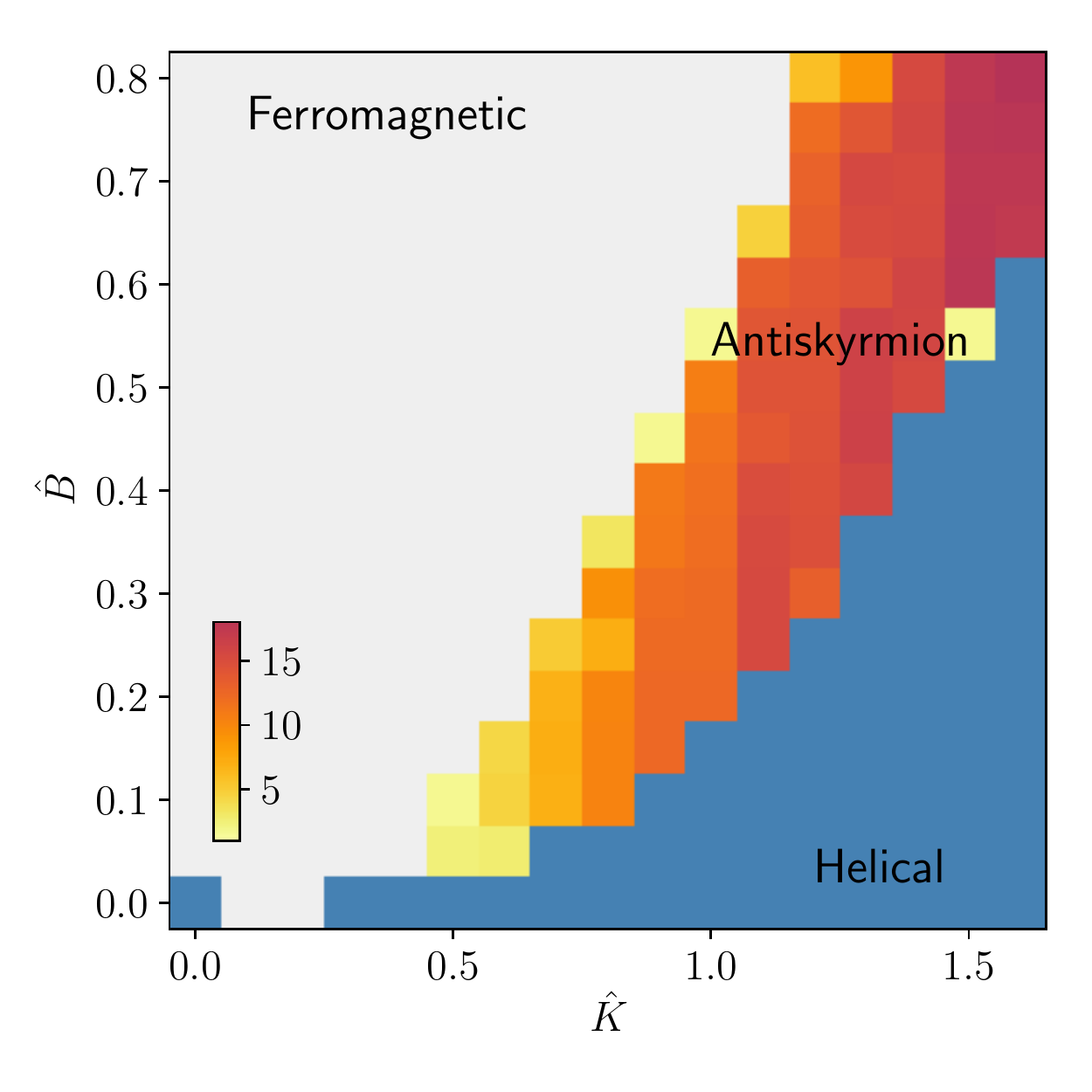}
 \caption{Antiskyrmion lattice phase within the DM interaction coefficient $\hat{K}$ and the magnetic field $\hat{B}$. Here we use a ZFC schedule to a fixed target temperature of $\hat{T} = 0.9$ everywhere. The color-coding illustrates the total antiskyrmion number, i.e.~the antiskyrmion phase is shown in red.}
 \label{fig:K-phases}
\end{figure}

In Fig.~\ref{fig:K-phases}, we show the corresponding phase diagram for $\hat{B}$ and $\hat{K}$, using an experimentally motivated ZFC schedule.
We observe that larger values of $\hat{K}$ require a larger magnetic field in order for antiskyrmion tubes to form.
Since materials with a large DM parameter are rare in practice, we note that experiments might look for antiskyrmions in materials with reasonably low DM interaction strength using a comparably small magnetic field.
However, at the same time, this approach is limited by the fact that, if the magnetic field is too small, the antiskyrmion lattice phase disappears completely.
Physically, in this case, the positive energy contribution by the magnetic field interaction is too small, thereby failing to stabilize the solitons.
This sets a lower limit on the DM interaction strength to stabilize antiskyrmion tubes at $\hat{K} \gtrsim 0.4$.
Indeed, this value can be crucial in the choice of material in an experimental search for antiskyrmions in chiral magnets.

\begin{figure}
 \centering
 \includegraphics[width=0.8\columnwidth]{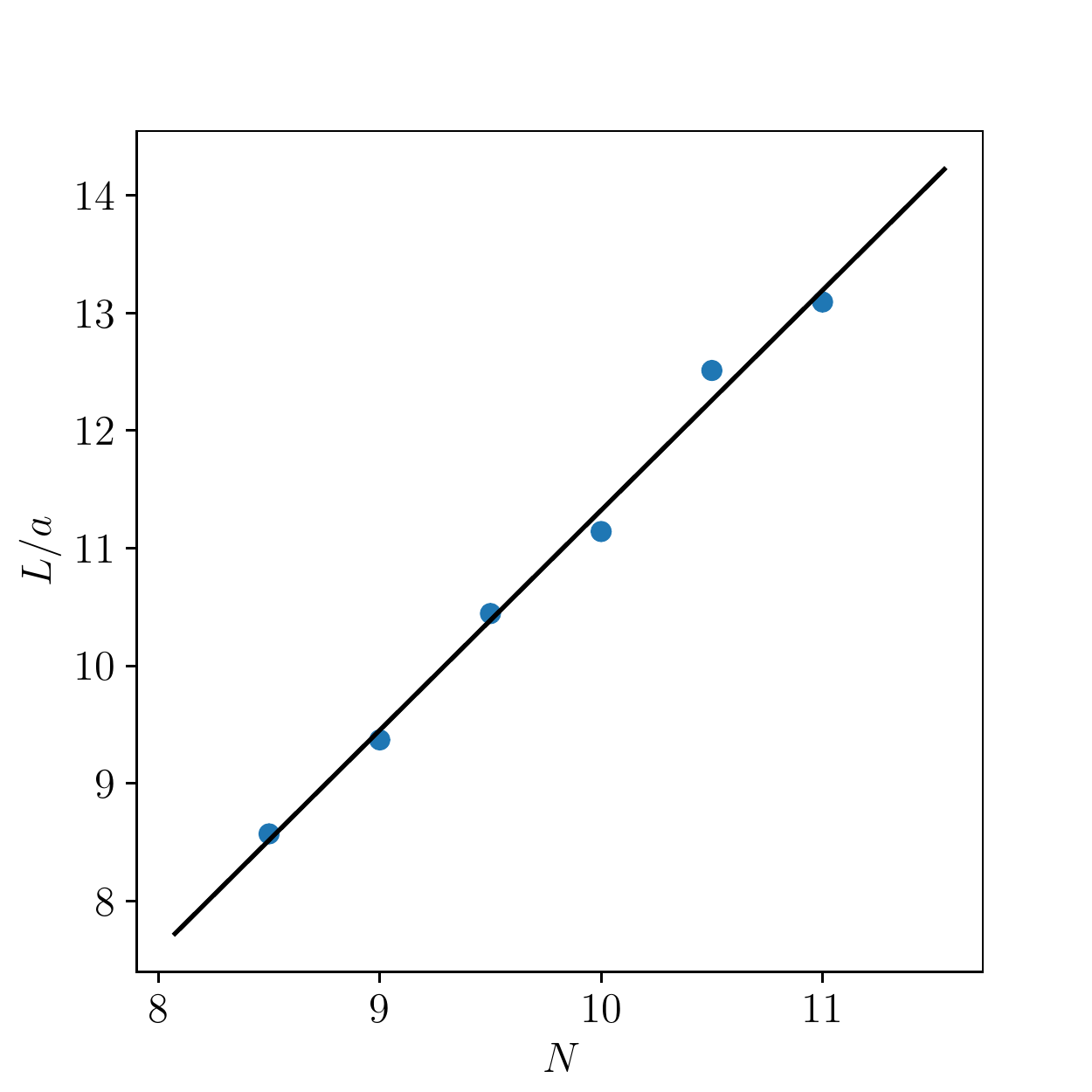}
 \caption{Average antiskyrmion size in units of the lattice spacing, $L / a$, as a function of the helical period in units of the lattice sites, $N$, (cf.~Eq.~\eqref{eq:N}). Here, we consider a target temperature and magnetic field of $\hat{T} = 0.9$ and $\hat{B} = 0.2$, respectively. These are obtained by performing a ZFC schedule on a $60 \times 60 \times 30$ spin lattice.}
 \label{fig:sizes}
\end{figure}

In addition, intuitively, the DM interaction strength controls the typical size of spin structures that emerge on the lattice.
Let us briefly quantify this explicitly, by defining the average antiskyrmion size as
\begin{equation}
	L \simeq \sqrt{\frac{A_{xy}}{Q_d}} \, .
\end{equation}
Here, $A_{xy}$ is the surface area of the lattice in the $xy$-plane and $Q_d$ is the total topological charge given in~\eqref{eq:top_charge_discrete}.
We count how many antiskyrmion tubes can be densely packed into any given lattice volume.
At the same time, in the helical phase, the period of the spin configurations (in terms of lattice sites) is given by~\cite{do2009skyrmions}
\begin{equation}
 N = \frac{2 \pi}{\arctan \hat{K}} \, .
 \label{eq:N}
\end{equation}
This is the defining estimate for the DM interaction coefficient that we used throughout this work for $N = 10$.
Intrinsically, the average antiskyrmion size should be related to the period of spin configurations.
Indeed, in Fig.~\ref{fig:sizes}, we show the average antiskyrmion size as a function of $N$, for constant temperature and magnetic field, $\hat{T} = 0.9$ and $\hat{B} = 0.2$, respectively.
In this simulation, we use a $60 \times 60 \times 30$ lattice to reduce the effects of a finite lattice size and apply a ZFC annealing schedule.
We find that, to good agreement, both quantities are related linearly,
\begin{equation}
 \frac{L}{a} \approx 1.87 N - 7.38 \, ,
\end{equation}
thereby supporting the fact that the DM interaction coefficient $\hat{K}$ is controlling the size of the antiskyrmion spin textures.
Here, again, $a$ denotes the lattice spacing.
Finally, we hope that this analysis can guide the experimental search for antiskyrmions by identifying suitable magnetic materials.
We note there have been some claims of antiskyrmions existing in both chiral and centrosymmetric materials.
However to date the objects observed do not seem to correspond to the antiskyrmions stabilized by DM interactions observed by us in this study.
We hope that the sizes and shapes and behavior we have found in this study may help to resolve potential ambiguities.
However it is also noted that we are using a restricted model Hamiltonian and our study cannot produce magnetic objects such as magnetic bubbles that rely on dipole interactions, which may have been observed in previous experimental studies.

\section{Conclusions}
\label{sec:conclusions}

Although skyrmions have been studied in great detail both theoretically and experimentally, in particular in condensed matter physics (for a recent overview see~\cite{back20202020}), antiskyrmions have not received the same attention so far.
In this work, we have demonstrated that antiskyrmions may indeed be found in bulk magnetic materials dominantly featuring a DM interaction that corresponds to a $D_{2d}$ crystal structure.
To study the existence of stable magnetic antiskyrmions, we have used MC techniques utilizing a simulated annealing process that have been proven to correctly reproduce experimental results related to the formation of magnetic skyrmion tubes~\cite{Buhrandt:2013uma}.

In particular, we have classified three different types of DM couplings, which, in combination with the ferromagnetic exchange interaction, give rise to Bloch skyrmions, N\'eel skyrmions and antiskyrmions.
For the latter case, we have presented, for the first time, a finite-temperature phase diagram of the spin textures that emerge at different temperatures and magnetic fields.
We find that a hexagonal lattice of antiskyrmion tubes is stabilized in a large region of parameter space.
At the same time, hysteresis effects can deform the antiskyrmion phase as we change the annealing procedure of the simulation, consistent with experimental observations in other materials~\cite{birch2019increased,sukhanov2020robust}.
Nevertheless, independent of the precise annealing schedule, we observe a stable antiskyrmion pocket around the parameters $\hat{T} = 0.9$ and $\hat{B} = 0.15$, strongly supporting the existence of magnetic antiskyrmion tubes in the material.

In addition, for fixed magnetic field and temperature, the range of values of the DM interaction strength that supports antiskyrmions is bounded from above and from below.
In particular, increasing the magnetic field rises both the lower and the upper phase boundary.
At the same time, at relatively low values of the magnetic field, the antiskyrmion phase disappears completely, which sets an absolute lower bound on the DM interaction strength of $\hat{K} \gtrsim 0.4$.
Therefore, to experimentally study the formation and stability of antiskyrmions in chiral magnets, a $D_{2d}$ material with a sufficient DM interaction strength is necessary.
In this context, our work can provide crucial experimental guidance in searching for stable antiskyrmions in magnetic materials.

In the future, we hope that these simulations can also shed light into the dynamics of antiskyrmion creation and annihilation as well as their interactions (similar to earlier studies~\cite{Foster:2019rbd,capic2020skyrmion,brearton2020magnetic,leonov2017asymmetric,Ross:2020hsw}).
As a proof of principle, we have indicated the existence of Bloch points mediating the phase transition towards a stable antiskyrmion lattice phase.
The dynamics of this process certainly merit further investigations.

\acknowledgments
This work was supported by the UK Skyrmion Project EPSRC Programme Grant (EP/N032128/1).
S.S.~is funded by the Deutsche Forschungsgemeinschaft (DFG, German Research Foundation) -- 444759442.

\bibliography{references,references_inspire}

\end{document}